\documentclass{mem}
\usepackage{natbib}\usepackage{txfonts}\usepackage{balance}
\usepackage{graphicx}
\usepackage[a4paper,breaklinks,dvipdfm]{hyperref}
\idline{000}{000}
\begin{document}
\def\teff{$T\rm_{eff }$}
\def\kms{$\mathrm {km s}^{-1}$}

\title{A bathtub model for the star-forming interstellar medium }

\subtitle{}

\author{ A. \,Burkert\inst{1} }

\institute{University Observatory Munich (USM), Scheinerstrasse 1, 81679 Munich,
German, \email{burkert@usm.lmu.de} }

\authorrunning{Burkert}

\titlerunning{ISM Bathtub}

\abstract{
The bathtub model of the star forming interstellar medium (ISM) is based on the powerful constraint that
mass has to be conserved when gas flows through its various thermal and density phases,
ending up eventually in a young star or being blown away by stellar feedback. It
predicts that the star formation rate of a molecular cloud is not determined by the cloud's mass or its
internal collapse timescale, but rather by the accretion rate of new gas. For the most simple
case of a constant accretion flow an equilibrium state is reached quickly where the star formation rate equals
the accretion rate and where the dense gas mass is constant and independent of time.
The mass of the young star cluster, on the other hand, increases linearly with time. The stellar
mass fraction therefore represents a sensitive clock to measure the age of the star-forming region.
The bathtub models predicts that the efficiency of star formation is small, of order 1\%, even in the dense
filamentary phases of molecular clouds. It provides a simple
explanation for the dense gas fraction of order 10\% in molecular clouds and large gas depletion timescales
of star-forming galaxies of order $5 \times 10^8 - 10^9$ yrs. 

}
\maketitle{}

\section{Introduction}

Stars form from cold and dense molecular gas.
The star formation rate (SFR), averaged over a whole galaxy, correlates well with
the total molecular gas mass M$_{H_2}$ with a surprisingly
universal depletion timescale $\tau_{depl} = M_{H_2}/SFR \approx  10^9$ yrs
(Tacconi+ 2017). 

On small scales, however, the physical properties of molecular clouds and as
a result their star formation histories vary strongly from place to place. Francesco Palla's work has
been pioneering in outlining the complexity of local star formation and in providing insight into
properties of star forming regions, like their typical star formation timescales which are of order $2 - 4 \times 10^6$ yrs
and their star formation histories (Palla \& Stahler 1999). In order to explain the inefficiency of star formation
Lada et al. (2010, see also Burkert \& Hartmann 2013) suggested that
there exists a density threshold of order $n \approx 10^4$ cm$^{-3}$ for star formation.
Lee \& Hennebelle (2016) termed this dense star-forming region within a molecular cloud
the gaseous protocluster. Interestingly, observations indicate that the mass fraction
of dense gas is rather universal with a value of 10$\%$ the total molecular cloud mass
(Lada et al. 2010; Vutisalchavakul et al. 2016). Even more interesting is the
fact that even in this dense gas phase the stellar mass fraction is small, of the
order of 10$\%$.

One serious observational caveat in unraveling the star formation history of a molecular cloud
is the fact that the presently observed young stellar population formed under physical conditions
that are not directly observable anymore. The current structure of the molecular cloud, on the other hand,
determines future star formation which is not yet observable. Only in an equilibrium state can the past
be directly related to the future.
One would probably at first focus on the available dense molecular gas as the primary site of star formation.
This concept is however misleading as molecular clouds and especially their dense gas regions are not isolated.
As shown below it is actually not the amount of dense gas or its gravitational collapse timescale that determines
the star formation rate, but rather the rate of dense gas formation.

Here I can only introduce the most simple version of the ISM
bathtub model (see also Bouche et al. 2010 for the galactic bathtub and Burkert \& Hartmann 2013). 
A detailed investigation of more complex situations, including gravity driven accretion and
exponentially increasing star formation rates will be discussed in a subsequent paper (Burkert, in preparation).

\section{The simple bathtub model (SBT) of the star-forming ISM}
Let us consider a star-forming region that is powered by a constant inflow of gas $\dot{M}_{acc}$.
In addition, we apply assumptions, often also used in numerical simulations, of a critical gas density threshold
$n_{dense} = 10^4$ cm$^{-3}$ (Lada et al. 2010), above which stars form
on a local free fall timescale (Krumholz et al. 2012) $\tau_{ff,dense} \approx 4 \times 10^5$ yrs
and a star formation rate that correlates linearly with the dense gas mass, multiplied by
a constant efficiency per free-fall time $\epsilon_{ff}$:

\begin{eqnarray}
\frac{d M_{dense}}{dt}  =   \dot{M}_{acc} - \frac{M_{dense}}{\tau_{ff,dense}}\\
SFR  =   \frac{dM_*}{dt}  =   \epsilon_{ff} \frac{M_{dense}}{\tau_{ff,dense}} 
\end{eqnarray}

\noindent Without gas accretion, i.e. $\dot{M}_{acc} = 0$, the solution is
$M_{dense} = M_0 \times \exp (-t/\tau_{ff,dense})$ and the dense gas mass and SFR decreases exponentially with time.
This solution suffers from the fact that
there is no explanation how the initial dense gas reservoir M$_0$ was generated and why star formation did
not already start during that phase. In addition,
as shown e.g. by Palla \& Stahler (1999), the star formation rates
in some star-forming region are actually {\it increasing} exponentially with time
which is just the opposite of the solution that we get without accretion.

Including a non-zero accretion term the solution is fundamentally different:
\begin{eqnarray}
M_{dense} & = & \dot{M}_{acc} \tau_{ff,dense} \left[1 - \exp \left(-\frac{t}{\tau_{ff,dense}} \right) \right] \nonumber \\
SFR & = & \epsilon_{ff} \dot{M}_{acc} \left[1 - \exp \left(-\frac{t}{\tau_{ff,dense}} \right) \right] 
\end{eqnarray}
\noindent We can now identify two phases. For t$ < \tau_{ff,dense}$ the dense gaseous protocluster mass increases linearly with
time $M_{dense} = \dot{M}_{acc} \times t$, independent of $\tau_{ff,dense}$,  as does the SFR. This is the non-equilibrium filling phase of the bathtub.
The situation changes for t$ >\tau_{ff,dense}$. The system now approaches a constant dense gas mass 
$M_{dense} = \dot{M}_{acc} \times \tau_{ff,dense}$ and a constant star formation rate
$SFR = \epsilon_{ff} \times \dot{M}_{acc}$, despite the fact that gas is still continuously being accreted. 
In this equilibrium phase, the SFR does not depend at all on the amount of dense gas M$_{dense}$,
nor on the physics of star formation, hidden in $\tau_{ff,dense}$. It however depends critically on the physics of accretion.
This demonstrates that it is the flow of gas through
the various gas phases of the ISM that determines its star formation rate as function of time
and not the instantaneous amount of
dense gas that is observed at any given time. Indeed, being constant and time independent, the dense gas mass does not
contain any information about the past or future star formation history.
Most observed star forming regions have ages of order a few Myrs, which is much longer than $\tau_{ff,dense}$.
They should therefore be in this equilibrium phase.

\section{The diffuse molecular gas reservoir and the dense gas fraction}

Despite being the primary site of star formation, the dense molecular gas phase
represents only a small fraction of the galactic molecular gas and is surrounded by a
diffuse molecular gas reservoir with densities of order n$_{diff} \approx 10^2$ cm$^{-3}$
and corresponding free fall timescales of $\tau_{diff} \approx 4 \times 10^6$ yrs.
It has long been assumed that this diffuse component is supported against gravitational collapse by internal turbulent flows.
New observations and theoretical models however indicate that the observed velocities are
actually signatures of gravitational collapse onto the embedded dense, star-forming gas regions
(e.g.  Ballesteros-Paredes 2011).
In this case, the diffuse component is an ideal candidate for the gas reservoir that feeds the dense gas
(Mac Low et al. 2017). Let us now investigate the implication of this scenario within the SBT model. 
We approximate the infall rate of diffuse gas by $\dot{M}_{acc,dense}$ = M$_{diff}/\tau_{ff,diff}$. In the equilibrium phase, 
discussed above, we then find

\begin{eqnarray}
M_{dense} & = &  \dot{M}_{acc,dense} \tau_{ff,dense} =\\
  & = &  M_{diff} \times \frac{\tau_{ff,dense}}{\tau_{ff,diff}} = 0.1 M_{diff} \nonumber
\end{eqnarray}

\noindent which is in excellent agreement with observations. Note that the dense gas fraction
depends critically on the ratio of the collapse timescales of the diffuse and dense gas component which could
vary strongly with galactic environment.

\section{Origin of the large gas depletion timescale in galaxies}

In the SBT equilibrium phase the dense gas mass and the SFR is constant. 
The stellar mass is however linearly increasing with time 

\begin{equation}
M_* \sim \epsilon_{ff} \dot{M}_{acc} \times t. 
\end{equation}

\noindent $M_*/M_{dense}$ therefore is very sensitive to the evolutionary time t of the star forming region 
and in late phases could in principle approach values of unity or larger. 
For typical local star-forming regions in the solar neighborhood
M$_* \approx 0.1$ M$_{dense}$. It can be shown (Burkert, in preparation, see Fig. 1)
that for typical ages of star forming regions t $\approx 10 \times \tau_{ff,dense}$
this implies that even in the dense gaseous protocluster environment star formation must be rather
inefficient with values of $\epsilon_{ff} \approx 0.01$. This now allows us to determine a molecular gas depletion
timescale. With M$_{H_2}$ = M$_{diff}$ + M$_{dense} \approx$ M$_{diff}$ and using the previous equations we find

\begin{eqnarray}
\tau_{depl} & = &  \frac{M_{H_2}}{SFR} = \frac{\tau_{ff,diff}}{\epsilon_{ff}} \\
 & \approx & 100 \times \tau_{ff,diff} \approx 4 \times 10^8 yrs \nonumber
\end{eqnarray}

\noindent In the framework of the SBT model it is therefore the combination of the relatively large 
collapse timescale of the diffuse molecular gas, combined with the low efficiency of star formation 
in the dense molecular gas component that determines the global gas depletion timescales of galaxies.
\begin{figure*}[t!]
\resizebox{\hsize}{!}{\includegraphics[clip=true]{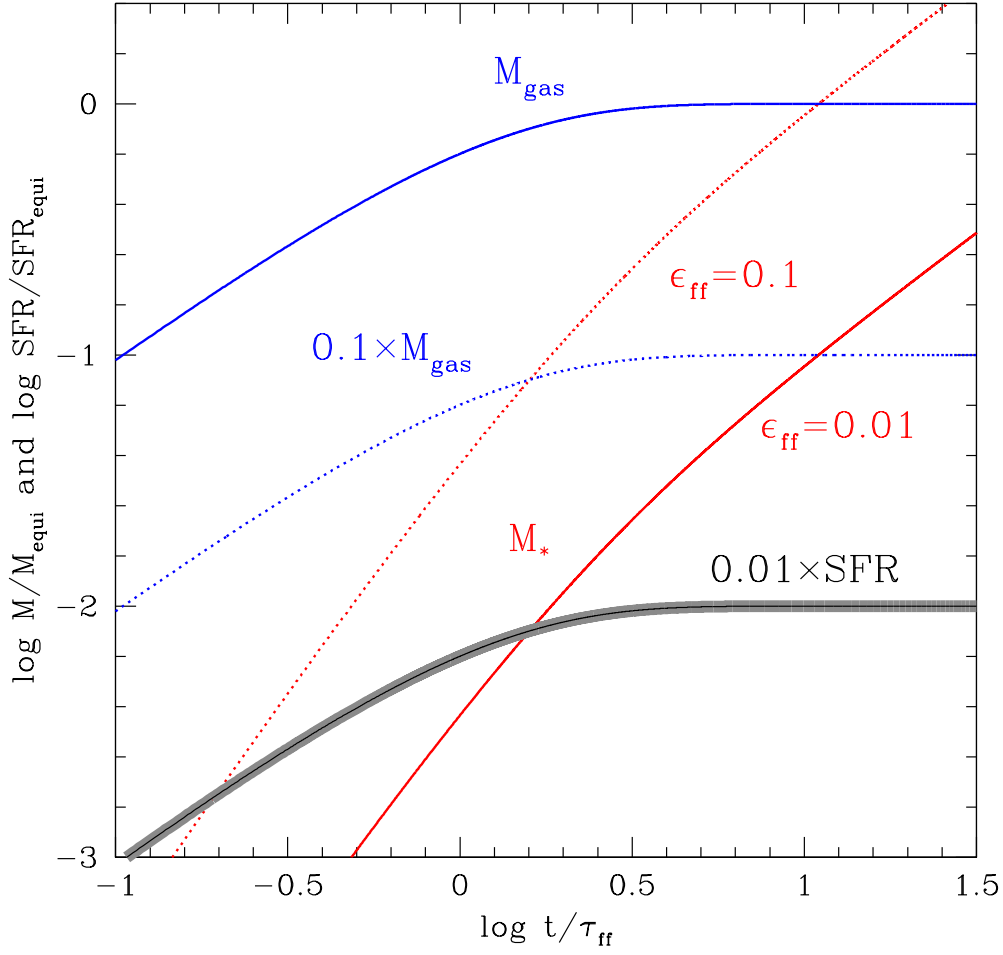}}
\caption{\footnotesize The evolution of the logarithm of the dense gas mass (blue solid line) and of
the stellar mass (red lines) in the gaseous protocluster is shown as function of time. 
The stellar mass is shown for star formation efficiencies $\epsilon_{ff}$ of 0.1 (dotted red line)
and for $\epsilon_{ff} = 0.01$ (solid red line).
Masses are scaled to the dense gas equilibrium mass $\dot{M}_{acc} \tau_{ff,dense}$. 
The time t is normalized to the free fall time $\tau_{ff,dense}$ of the dense gas component.
The grey thick line shows the star formation rate SFR, normalized to its equilibrium value
$\epsilon_{ff} \times \dot{M}_{acc}$. For better visibility, the SFR has been shifted to 0.01 its actual value.
The dotted blue line shows 10\% of the dense gas mass which is the typical mass of the  young stellar 
component, observed in nearby star-forming regions. One can see that for $\epsilon_{ff} = 0.01$
the stellar component reaches 10\% the gas mass at about 10 $\tau_{ff,dense}$, which is in agreement
with observations.  }
\label{eta}
\end{figure*}
\begin{acknowledgements}
I would like to thank Edvige Corbelli for taking care of me (as always) and the organisers for their invitation
to an inspiring conference in memory of Francesco Palla. I will miss his wise ideas
and friendly smile.
\end{acknowledgements}

\bibliographystyle{aa}

\end{document}